\pdfoutput=1
\RequirePackage{ifpdf}
\ifpdf 
\documentclass[pdftex]{sigma}
\else
\documentclass{sigma}
\fi

\usepackage{tikz}

\numberwithin{equation}{section}

\newtheorem{Theorem}{Theorem}[section]
{ \theoremstyle{definition}
	\newtheorem{Remark}[Theorem]{Remark}
	\newtheorem{Summary}[Theorem]{Summary}
}

\let\p\partial

\tikzstyle{Black vertex}=[fill=black, draw=black, shape=circle]
\tikzstyle{Empty vertex}=[fill=white, draw=black, shape=circle]
\tikzstyle{gray node}=[fill={rgb,255: red,191; green,191; blue,191}, draw=black, shape=circle]

\tikzstyle{plain}=[-]
\tikzstyle{directed}=[->, draw={rgb,255: red,128; green,128; blue,128}]
\tikzstyle{new edge style 0}=[-, dashed]

\begin{document}

\newcommand{\arXivNumber}{2101.04332}

\renewcommand{\PaperNumber}{108}

\FirstPageHeading

\ShortArticleName{The Lattice Sine-Gordon Equation as a Superposition Formula for an NLS-Type System}

\ArticleName{The Lattice Sine-Gordon Equation\\ as a Superposition Formula for an NLS-Type System}

\Author{Dmitry K.~DEMSKOI}

\AuthorNameForHeading{D.K.~Demskoi}

\Address{School of Computing, Mathematics and Engineering, Charles Sturt University,\\
NSW 2678, Australia}
\Email{\href{mailto:ddemskoy@csu.edu.au}{ddemskoy@csu.edu.au}}
\URLaddress{\url{http://mathassess.net/demskoy/}}

\ArticleDates{Received June 30, 2021, in final form December 13, 2021; Published online December 21, 2021}

\Abstract{We treat the lattice sine-Gordon equation and two of its generalised symmetries as a compatible system. Elimination of shifts from the two symmetries of the lattice sine-Gordon equation yields an integrable NLS-type system. An auto-B\"acklund transformation and a superposition formula for the NLS-type system is obtained by elimination of shifts from the lattice sine-Gordon equation and its down-shifted version. We use the obtained formulae to calculate a~superposition of two and three elementary solutions.}


\Classification{35Q51; 35Q55; 37K60}

\section{Introduction}
Integrable differential-difference equations (D$\Delta$Es) with one continuous and one discrete variable are known to be closely related with integrable
partial differential equations (PDEs). In particular, many integrable D$\Delta$Es may be interpreted as B\"acklund transformations of some PDEs \cite{levi}. For example, the integrable Volterra-type equations
\begin{gather}
	\frac{\p}{\p x}u_l=f(u_{l-1},u_l,u_{l+1}) \label{voltgen}
\end{gather}
are known to be related to the NLS-type PDEs \cite{shabyam} by means of elimination of shifts.
More precisely, if one considers (\ref{voltgen}) and its simplest generalised symmetry
\begin{gather}
	\frac{\p}{\p t}u_l=g(u_{l-2},u_{l-1},u_l,u_{l+1},u_{l+2}),
	\label{voltsym}
\end{gather}
then elimination of $u_{l-2}$, $u_{l+1}$, $u_{l+2}$ from the system of (\ref{voltgen}) and (\ref{voltsym}) yields a two component system of the NLS-type on the quantities $u_{l-1}$ and $u_{l}.$ A by-product of this calculation is an invertible auto-transformation of the resulting system of PDEs generated by~(\ref{voltgen}).

As far as construction of exact solutions is concerned, a more important class of transformations is
auto-B\"acklund transformations with a spectral parameter which are usually non-invertible.
A direct calculation of such transformation is a tedious task. The knowledge of other
structures associated with integrability, e.g., a Lax pair or Painlev\'e structure, may significantly
speed up the search for such a transformation~\cite{weiss}.

In \cite{demsknls} we showed that a system obtained by elimination of shifts from (\ref{voltgen}) and (\ref{voltsym}), where at the same time equations (\ref{voltgen}) and (\ref{voltsym}) represent symmetries of a quad-equation, automatically possesses an auto-B\"acklund transformation with spectral parameter. The method was illustrated with using the lattice $H_1$ equation \cite{ABS} (lattice potential KdV equation) and a pair of its simplest symmetries.

In this paper we use the method of \cite{demsknls} to find an auto-B\"acklund transformation and the superposition
formula for solutions of the system{\samepage
\begin{gather}
\begin{split}
&p_t= -p_{xx} + q_x p_x^2-q_x, \\
&q_t= q_{xx} + p_x q_x^2-p_x
\end{split}
	\label{NLSd}
\end{gather}
using its connection with the lattice sine-Gordon equation and the Volterra-type equations.}

System (\ref{NLSd}) is reminiscent of the well known derivatives NLS system, yet it belongs to a~different cluster of NLS-type systems labelled as (d) in~\cite{mikh}.
All systems in the cluster are related by transformations that preserve local conservation laws.
Therefore, having a procedure for generating solutions for~(\ref{NLSd}) allows one to generate solutions for other systems in the cluster as well.

\subsection{Symmetries of the lattice sine-Gordon equation}
\label{overview}
We consider the lattice sine-Gordon (lsG) equation
\begin{gather}
 \alpha (u_{\scriptscriptstyle l,m}u_{\scriptscriptstyle l+1,m+1}-u_{\scriptscriptstyle l+1,m}u_{\scriptscriptstyle l,m+1})+\beta u_{\scriptscriptstyle l,m}u_{\scriptscriptstyle l+1,m}u_{\scriptscriptstyle l,m+1}u_{\scriptscriptstyle l+1,m+1}=1,	\label{lSGgen}
\end{gather}
where $\alpha$ and $\beta$ are arbitrary parameters. It was first derived by Hirota in a slightly different form by the method of dependent variable
transformation \cite{hirotsg}. The unknown function $u$ is assumed to depend on the three continuous variables~$t$,~$x$, and~$y$ as well as two discrete variables $l$ and $m$. The dependence on the continuous variables is determined through the {\it generalised symmetries} of equation~(\ref{lSGgen}) while discrete variable $l$ labels the components of the symmetries, and~$m$ counts iterations of the auto-B\"acklund transformation.

Equation (\ref{lSGgen}) degenerates into the lattice Liouville equation \cite{hirotaliouv,tremblay}
\begin{gather}
	\alpha (u_{\scriptscriptstyle l,m}u_{\scriptscriptstyle l+1,m+1}-u_{\scriptscriptstyle l+1,m}u_{\scriptscriptstyle l,m+1})=1
	\label{latliouv}
\end{gather}
when $\beta=0$. This connection turns out to be useful for the construction of symmetries of (\ref{lSGgen}). First, we point out that equation (\ref{latliouv}) is {\em Darboux integrable}. That is, it possesses integrals with respect to both discrete variables. The integrals
\begin{gather}
	w_{\scriptscriptstyle l,m}=\frac{u_{\scriptscriptstyle l,m}}{u_{\scriptscriptstyle l+1,m}+u_{\scriptscriptstyle l-1,m}},\qquad \bar{w}_{\scriptscriptstyle l,m}=\frac{u_{\scriptscriptstyle l,m}}{u_{\scriptscriptstyle l,m+1}+u_{\scriptscriptstyle l,m-1}},
	\label{ints}
\end{gather}
satisfy the relations
\[(S_m-1)w_{\scriptscriptstyle l,m}=0,\qquad (S_l-1)\bar{w}_{\scriptscriptstyle l,m}=0,\]
on solutions of (\ref{latliouv}). Here $S_l$ and $S_m$ stand for the respective shift operators, e.g., $S_l(w_{\scriptscriptstyle l,m})=w_{\scriptscriptstyle l+1,m}$. These integrals can be easily derived from the determinantal structure of (\ref{latliouv}) (see, e.g.,~\cite{demtran}).

Then, the construction of the higher symmetries of (\ref{latliouv}) is rather algorithmic due to a method presented in
\cite{adst} which allows one to construct operators that map the integrals of a Darboux integrable equation to its symmetries. Since the lattice Liouville equation is Darboux integrable, the structure of its symmetries is determined by such operators, i.e., the symmetries can be cast into either of the forms
\begin{gather}
\p_x u_{\scriptscriptstyle l,m}=M F(w),\qquad \p_x u_{\scriptscriptstyle l,m}=\bar{M} F(\bar{w}),
	\label{latLiouvsym}
\end{gather}
where $w$ and $\bar{w}$ are the integrals of (\ref{latliouv}) or constants.
Skipping technical details, we present the operators explicitly:
\[M=w_{\scriptscriptstyle l,m} (u_{\scriptscriptstyle l-1,m} S_l-u_{\scriptscriptstyle l+1,m}), \qquad
\bar{M}=\bar{w}_{\scriptscriptstyle l,m} (u_{\scriptscriptstyle l,m-1} S_m-u_{\scriptscriptstyle l,m+1}).
\]
Since equation (\ref{latliouv}) is invariant with respect to the involution $u_{\scriptscriptstyle l\pm i,m}\to u_{\scriptscriptstyle l\mp i,m}$, there exists another set of operators
\begin{gather*} {\cal M}=w_{\scriptscriptstyle l,m}\big(u_{\scriptscriptstyle l+1,m} S_l^{-1}-u_{\scriptscriptstyle l-1,m}\big), \\
\bar{\cal M}=\bar{w}_{\scriptscriptstyle l,m} \big(u_{\scriptscriptstyle l,m+1} S_m^{-1}-u_{\scriptscriptstyle l,m-1}\big)
\end{gather*}
with the same property, that is ${\cal M}(F(w))$ and $\bar{\cal M}(F(\bar{w}))$ are the symmetries of (\ref{latliouv}). Obviously all the statements regarding the integrals and symmetries of~(\ref{lSGgen}) and~(\ref{latliouv}) involve two identical parts: one involving variables $u_{\scriptscriptstyle l+i,m}$ and the other variables $u_{\scriptscriptstyle l,m+i}$. Therefore, for convenience, the second part will be omitted whenever it is not essential.

Equations (\ref{latLiouvsym}) are too general to be all integrable. Nevertheless, for some choices of function~$F$ this is the case, in particular, when~$F$ is chosen such that~(\ref{latLiouvsym}) is also a symmetry of~(\ref{lSGgen}). The simplest choice $F=1$ delivers the equation
\begin{gather}
	\p_x u_{\scriptscriptstyle l,m}=\frac{u_{\scriptscriptstyle l,m} (u_{\scriptscriptstyle l-1,m}-u_{\scriptscriptstyle l+1,m})}{u_{\scriptscriptstyle l-1,m}+u_{\scriptscriptstyle l+1,m}},
	\label{sym1}
\end{gather}
which is the simplest generalised symmetry of (\ref{lSGgen}). That is, on solutions of (\ref{lSGgen}) and (\ref{sym1}) the relation
\[\p_x\left(\alpha (u_{\scriptscriptstyle l,m}u_{\scriptscriptstyle l+1,m+1}-u_{\scriptscriptstyle l+1,m}u_{\scriptscriptstyle l,m+1})+\beta u_{\scriptscriptstyle l,m}u_{\scriptscriptstyle l+1,m}u_{\scriptscriptstyle l,m+1}u_{\scriptscriptstyle l+1,m+1}\right)=0\]
is satisfied identically.

Further, the fact that (\ref{sym1}) is a symmetry of (\ref{latliouv}) implies that on solutions of these equations the derivative $\p_x$ commutes with the shift $S_l$, which in turn, implies that $\p_x w_{\scriptscriptstyle l,m}$ can be expressed in terms of $w_{\scriptscriptstyle l,m}$ itself and its shifts. Therefore the integral $w_{\scriptscriptstyle l,m}$ provides us with a Miura-type transformation from equation $(\ref{sym1})$ into the modified Volterra equation
\begin{gather}
	\p_x w_{\scriptscriptstyle l,m}=2 w_{\scriptscriptstyle l,m}^2 (w_{\scriptscriptstyle l-1,m}-w_{\scriptscriptstyle l+1,m} ).
	\label{mvolterra}
\end{gather}
The complete classification of the Volterra-type equations can be found in \cite{yam06}.
As it is pointed out in \cite{cherdyam}, the Volterra-type equations possess {\it local} master symmetries.
For example, a master symmetry of (\ref{mvolterra}) is the equation
\begin{gather}
	\p_y w_{\scriptscriptstyle l,m}=l\p_x w_{\scriptscriptstyle l,m}+w_{\scriptscriptstyle l,m}^2 (w_{\scriptscriptstyle l-1,m}+w_{\scriptscriptstyle l+1,m} ).
	\label{msmvolterra}
\end{gather}
The commutator $(\p_y\p_x -\p_x\p_y)w_{\scriptscriptstyle l,m}$, calculated on solutions of (\ref{mvolterra}) and (\ref{msmvolterra}) gives the simplest generalised symmetry of (\ref{mvolterra}).
The locality of master symmetries significantly simplifies their construction. Moreover, it has been observed that for the Volterra-type equations the master symmetries bear resemblance of the equation itself. The present case is no exception: a master symmetry of~(\ref{sym1}) is given by
$\p_y u_{\scriptscriptstyle l,m}=l\p_x u_{\scriptscriptstyle l,m}$.
Hence the simplest generalised symmetry of~(\ref{sym1}) has the form
\[\p_t u_{\scriptscriptstyle l,m}=(\p_y\p_x -\p_x\p_y)u_{\scriptscriptstyle l,m},\]
which is explicitly given by the equation
\begin{gather}
	\p_t u_{\scriptscriptstyle l,m}= \frac{4u_{\scriptscriptstyle l-1,m} u_{\scriptscriptstyle l,m}^2 u_{\scriptscriptstyle l+1,m} (u_{\scriptscriptstyle l+2,m}-u_{\scriptscriptstyle l-2,m})}{(u_{\scriptscriptstyle l-2,m}+u_{\scriptscriptstyle l,m})(u_{\scriptscriptstyle l-1,m}+u_{\scriptscriptstyle l+1,m})^2(u_{\scriptscriptstyle l+2,m}+u_{\scriptscriptstyle l,m})}.
	\label{sym2}
\end{gather}
Both equations (\ref{sym1}) and (\ref{sym2}) are the symmetries of (\ref{lSGgen}).
So, the remainder of the paper focusses on a system of equations comprising of (\ref{sym1}), (\ref{sym2}), and (\ref{lSGgen}). We now assume that $\beta\ne 0$ hence by rescaling of the dependent variable we can set $\beta=1$:
\begin{gather}
	\begin{array}{l}
		\alpha (u_{\scriptscriptstyle l,m}u_{\scriptscriptstyle l+1,m+1}-u_{\scriptscriptstyle l+1,m}u_{\scriptscriptstyle l,m+1})+u_{\scriptscriptstyle l,m}u_{\scriptscriptstyle l+1,m}u_{\scriptscriptstyle l,m+1}u_{\scriptscriptstyle l+1,m+1}=1.
	\end{array}
	\label{lSG}
\end{gather}
\begin{Remark}In the preceding calculations, we derived the symmetries (\ref{sym1}) and (\ref{sym2}) in a way that highlights the connection with the lattice Liouville equation and its integrals. One could employ different approaches, e.g., calculate them directly or extract from previously published articles \cite{fxi1,fxi2}.
Also, it is worth mentioning that the integrals (\ref{ints}) are related to the conserved densities of equations (\ref{sym1}) and (\ref{sym2}), i.e., $\p_x \ln w_{l,m}, \p_t \ln w_{l,m} \in \mbox{Im}(S_l-1)$, which is in parallel with the continuous case. The connection between conserved densities and discrete Miura-type transformations has been noted before, see, e.g., \cite{adler2}.
\end{Remark}

\subsection{Connection with NLS-type systems}
As we already mentioned, the compatibility of a Volterra-type equation with its generalised symmetry yields an NLS-type system through elimination of shifts~\cite{shabyam}. In the present case,
 if we express the variables $u_{\scriptscriptstyle l+1,m}$, $u_{\scriptscriptstyle l+2,m}$, and $u_{\scriptscriptstyle l-2,m}$ from equation~(\ref{sym1}) to obtain
\begin{gather}
\begin{split}
& u_{\scriptscriptstyle l+1,m}=\frac{u_{\scriptscriptstyle l-1,m}(u_{\scriptscriptstyle l,m}-\p_x u_{\scriptscriptstyle l,m})}{u_{\scriptscriptstyle l,m}+\p_x u_{\scriptscriptstyle l,m}},\qquad u_{\scriptscriptstyle l+2,m}=\frac{u_{\scriptscriptstyle l,m}(u_{\scriptscriptstyle l+1,m}-\p_x u_{\scriptscriptstyle l+1,m})}{u_{\scriptscriptstyle l+1,m}+\p_x u_{\scriptscriptstyle l+1,m}},\\
& u_{\scriptscriptstyle l-2,m} = \frac{u_{\scriptscriptstyle lm}(u_{\scriptscriptstyle l-1,m}+\p_x u_{\scriptscriptstyle l-1,m})}{u_{\scriptscriptstyle l-1,m}-\p_x u_{\scriptscriptstyle l-1,m}}
\end{split}
	\label{shifts}
\end{gather}
and substitute them into (\ref{sym2}), we obtain a system of two equations on $u_{\scriptscriptstyle l-1,m}$ and $u_{\scriptscriptstyle l,m}$. Note that substitution of the expression for $u_{\scriptscriptstyle l+1,m}$ into the one for $u_{\scriptscriptstyle l+2,m}$ produces quite a cumbersome formula. Nevertheless, the resulting system turns out to be quite compact:
\begin{gather}
	\begin{split}
& \p_t u_{\scriptscriptstyle l-1,m} = -\p_x^2 u_{\scriptscriptstyle l-1,m}+\frac{(\p_x u_{\scriptscriptstyle l-1,m})^2 }{u_{\scriptscriptstyle l-1,m}}\left(\frac{\p_x u_{\scriptscriptstyle l,m}}{u_{\scriptscriptstyle l,m}}+1\right) -\frac{u_{\scriptscriptstyle l-1,m} \p_x u_{\scriptscriptstyle l,m}}{u_{\scriptscriptstyle l,m}},
\\
&	 \p_t u_{\scriptscriptstyle l,m} = \p_x^2 u_{\scriptscriptstyle l,m} + \frac{ (\p_x u_{\scriptscriptstyle l,m})^2 } {u_{\scriptscriptstyle l,m}}\left(\frac{\p_x u_{\scriptscriptstyle l-1,m}}{ u_{\scriptscriptstyle l-1,m}}-1\right) -\frac{u_{\scriptscriptstyle l,m} \p_x u_{\scriptscriptstyle l-1,m}}{u_{\scriptscriptstyle l-1,m}}.
\end{split}\label{NLS}
\end{gather}
Since both pairs $(u_{\scriptscriptstyle l-1,m},u_{\scriptscriptstyle l,m})$ and $(u_{\scriptscriptstyle l,m},u_{\scriptscriptstyle l+1,m})$ satisfy the same system, then (\ref{NLS}) admits an auto-transformation given explicitly by
\begin{equation*}
 (u_{\scriptscriptstyle l-1,m},u_{\scriptscriptstyle l,m} )\to
	\left(u_{\scriptscriptstyle l,m}, \frac{u_{\scriptscriptstyle l-1,m}(u_{\scriptscriptstyle l,m}-\p_x u_{\scriptscriptstyle l,m})}{u_{\scriptscriptstyle l,m}+\p_x u_{\scriptscriptstyle l,m}}\right).
\end{equation*}
System (\ref{NLS}) takes a particularly simple form (\ref{NLSd})
in the variables
\[u_{\scriptscriptstyle l-1,m}={\rm e}^{p},\qquad u_{\scriptscriptstyle l,m}={\rm e}^q.\]

As is the case with equations (\ref{sym1}) and (\ref{sym2}), the integral $w_{\scriptscriptstyle l,m}$ yields a Miura-type transformation between the respective NLS-type systems. More precisely, if we eliminate variables~$u_{\scriptscriptstyle l-2,m}$ and~$u_{\scriptscriptstyle l+1,m}$ from
the expressions for $w_{\scriptscriptstyle l-1,m}$ and $w_{\scriptscriptstyle l,m}$ by means of (\ref{shifts}), we obtain
\begin{gather}
	w_{\scriptscriptstyle l-1,m} = \frac12\frac{u_{\scriptscriptstyle l-1,m}-\p_x u_{\scriptscriptstyle l-1,m}}{u_{\scriptscriptstyle l,m}},\qquad w_{\scriptscriptstyle l,m} = \frac12\frac{u_{\scriptscriptstyle l,m}+\p_x u_{\scriptscriptstyle l,m}}{u_{\scriptscriptstyle l-1,m}}.
	\label{nlsmiur}
\end{gather}
Substitution (\ref{nlsmiur}) transforms (\ref{NLS}) into the combination of the derivative NLS system and its translational symmetry:
\begin{gather*}
\p_t w_{\scriptscriptstyle l-1,m}=-\p_x^2 w_{\scriptscriptstyle l-1,m}-4\p_x\big( w_{\scriptscriptstyle l-1,m}^2 w_{\scriptscriptstyle l,m}\big)+2\p_x w_{\scriptscriptstyle l-1,m},\\
\p_t w_{\scriptscriptstyle l,m}=\p_x^2 w_{\scriptscriptstyle l,m}-4\p_x\big( w_{\scriptscriptstyle l,m}^2 w_{\scriptscriptstyle l-1,m}\big)+2 \p_x w_{\scriptscriptstyle l,m}.
\end{gather*}
Therefore, the auto-B\"acklund transformation and superposition formulae obtained in the following section can also be used to build solutions of the derivative NLS, but not the other way around, as transformation~(\ref{nlsmiur}) is not invertible. We will refer to system~(\ref{NLS}) as a modified derivative NLS (mdNLS) system.

\section{Auto-B\"acklund transformation and superposition formula}\label{autobacklund}

Here we apply the method of \cite{demsknls} to derive exact solutions of (\ref{NLS}). The idea of using (\ref{lSG}) as a~superposition formula for solutions of (\ref{NLS}) is based on the fact that the symmetries of (\ref{lSG}) only depend on shifts with respect to one of the variables. So, equation (\ref{lSG}) possesses another set of symmetries of the form (\ref{sym1}) and (\ref{sym2})
with the substitution $u_{\scriptscriptstyle l+i,m}\to u_{\scriptscriptstyle l+i,m+1}$
applied. Elimination of shifts in this system, again, produces a system of the form (\ref{NLS}).
The process of constructing a solution consists of the following steps.

Starting with a seed solution $(u_{\scriptscriptstyle l-1,m},u_{\scriptscriptstyle l,m}),$ which often is a trivial solution, we would like to calculate another solution, labelled as $(u_{\scriptscriptstyle l-1,m+1},u_{\scriptscriptstyle l,m+1})$.
Let us consider the down-shifted, with respect to $l$, version of equation (\ref{lSG}):
\begin{gather}
 \alpha (u_{\scriptscriptstyle l-1,m}u_{\scriptscriptstyle l,m+1}-u_{\scriptscriptstyle l,m}u_{\scriptscriptstyle l-1,m+1})+u_{\scriptscriptstyle l-1,m}u_{\scriptscriptstyle l,m}u_{\scriptscriptstyle l-1,m+1}u_{\scriptscriptstyle l,m+1}=1.
	\label{2lSG}
\end{gather}
Equations (\ref{lSG}), (\ref{NLS}), and (\ref{2lSG}) can be schematically depicted as

\begin{center}
\begin{tikzpicture}[thick,scale=0.7, every node/.style={scale=0.8}]
 	\node [style=Empty vertex] (1) at (-16, 3) {};
		\node [style=Black vertex] (6) at (-16, 7) {};
		\node (19) at (-10, 7) {};
		\node (21) at (-21, 7) {};
		\node (22) at (-21, 3) {};
		\node (25) at (-20.75, 3.25) {${ u_{ l-1,m}}$};
		\node (31) at (-21, 6.75) {${ u_{ l-1,m+1}}$};
		\node (32) at (-15.25, 6.75) {${\hspace{2mm} u_{ l,m+1}}$};
		\node (33) at (-15.5, 3.25) {${ u_{ l,m}}$};
		\node (34) at (-11.25, 3.25) {${ u_{ l+1,m}}$};
		\node (35) at (-11, 6.75) {${\hspace{2mm}u_{ l+1,m+1}}$};
		\node (45) at (-10, 3) {};
		\node (46) at (-20, 8) {};
		\node (47) at (-16, 8) {};
		\node (48) at (-12, 8) {};
		\node (49) at (-12, 2) {};
		\node (50) at (-16, 2) {};
		\node (51) at (-20, 2) {};
		\node [style=gray node] (52) at (-12, 3) {};
		\node [style=gray node] (53) at (-12, 7) {};
		\node [style=gray node] (54) at (-20, 7) {};
		\node [style=Empty vertex] (56) at (-20, 3) {};
		\node (57) at (-18, 5) {lsG};
		\node (58) at (-14, 5) {lsG};
		\node (59) at (-18, 7.5) {mdNLS};
		\node (60) at (-14, 7.5) {mdNLS};
		\node (61) at (-18, 2.5) {mdNLS};
		\node (62) at (-14, 2.5) {mdNLS};
		\draw (1) to (6);
		\draw [style=new edge style 0] (1) to (50.center);
		\draw [style=new edge style 0] (6) to (47.center);
		\draw (54) to (6);
		\draw [style=new edge style 0] (21.center) to (54);
		\draw [style=new edge style 0] (54) to (46.center);
		\draw (1) to (52);
		\draw [style=new edge style 0] (52) to (45.center);
		\draw [style=new edge style 0] (52) to (49.center);
		\draw (6) to (53);
		\draw [style=new edge style 0] (53) to (19.center);
		\draw [style=new edge style 0] (53) to (48.center);
		\draw (52) to (53);
		\draw (56) to (1);
		\draw [style=new edge style 0] (22.center) to (56);
		\draw [style=new edge style 0] (56) to (51.center);
		\draw (56) to (54);
\end{tikzpicture}
\end{center}

\noindent In this diagram the white vertices represent a seed solution, the grey vertices represent the quantities that can be eliminated from the equations, and the black vertex represents the quantity that needs to be calculated.
In more detail, the rightmost vertices $u_{\scriptscriptstyle l+1,m}$ and $u_{\scriptscriptstyle l+1,m+1}$ can be eliminated from $(\ref{lSG})$ by means of $(\ref{shifts})_1$. This gives an expression containing the $x$-derivatives of $u_{\scriptscriptstyle l,m}$ and $u_{\scriptscriptstyle l,m+1}$.
Further, expressing $u_{\scriptscriptstyle l-1,m+1}$ from $(\ref{2lSG})$ and substituting in $(\ref{lSG})$ we obtain an equation on a single quantity $u_{\scriptscriptstyle l,m+1}$ which determines the $x$-dynamics of the solution. The resulting equation reads
\begin{gather}
	\big(\alpha^2-1\big)\p_x u_{\scriptscriptstyle l,m+1}=-A_{\scriptscriptstyle l,m} \p_x u_{\scriptscriptstyle l,m}+\alpha B_{\scriptscriptstyle l,m},
\label{xderivative}
\end{gather}
where
\[
A_{\scriptscriptstyle l,m}=\frac{(u_{\scriptscriptstyle l-1,m} u_{\scriptscriptstyle l,m+1}-\alpha) (\alpha u_{\scriptscriptstyle l,m+1} u_{\scriptscriptstyle l-1,m} -1) }{u_{\scriptscriptstyle l,m} u_{\scriptscriptstyle l-1,m}},\qquad B_{\scriptscriptstyle l,m}=\frac{ u_{\scriptscriptstyle l-1,m}^2 u_{\scriptscriptstyle l,m+1}^2-1}{ u_{\scriptscriptstyle l-1,m}}.
\]
The $t$-dependence of $u_{\scriptscriptstyle l,m+1}$ is determined by eliminating the $x$-derivatives in the respective mdNLS system by means of (\ref{xderivative}). This yields the following equation
\begin{gather}
 \big(\alpha^2-1\big)\p_t u_{\scriptscriptstyle l,m+1}
	=-A_{\scriptscriptstyle l,m}\p_x^2 u_{\scriptscriptstyle l,m} + \frac{A_{\scriptscriptstyle l,m}(u_{\scriptscriptstyle l-1,m}-\p_x u_{\scriptscriptstyle l-1,m}) (\p_x u_{\scriptscriptstyle l,m})^2}{u_{\scriptscriptstyle l,m} u_{\scriptscriptstyle l-1,m}}
\nonumber\\
\hphantom{\big(\alpha^2-1\big)\p_t u_{\scriptscriptstyle l,m+1}=}{}
+\frac{\alpha B_{\scriptscriptstyle l,m} \p_x u_{\scriptscriptstyle l,m} \p_x u_{\scriptscriptstyle l-1,m}}{u_{\scriptscriptstyle l,m}u_{\scriptscriptstyle l-1,m}}
	-\frac{\big(\alpha^2+1\big)u_{\scriptscriptstyle l,m+1}\p_x u_{\scriptscriptstyle l-1,m}}{u_{\scriptscriptstyle l-1,m}}
\nonumber\\
\hphantom{\big(\alpha^2-1\big)\p_t u_{\scriptscriptstyle l,m+1}=}{} +\frac{\alpha \big(\alpha^2+1\big) C_{\scriptscriptstyle l,m}\p_x u_{\scriptscriptstyle l,m}}{\big(\alpha^2-1\big) u_{\scriptscriptstyle l,m} }- \frac{4 \alpha^2 u_{\scriptscriptstyle l,m+1} \p_x u_{\scriptscriptstyle l,m}}{\big(\alpha^2-1\big) u_{\scriptscriptstyle l,m}} - \frac{\alpha\big(\alpha^2+1\big)}{\alpha^2-1}B_{\scriptscriptstyle l,m},
	\label{tderivative}
\end{gather}
where
\[C_{\scriptscriptstyle l,m}=\frac{1+u_{\scriptscriptstyle l-1,m}^2 u_{\scriptscriptstyle l,m+1}^2}{u_{\scriptscriptstyle l-1,m}}.\]

\begin{Remark}The requirement of compatibility of (\ref{xderivative}) and (\ref{tderivative}) yields the relation
\[
\frac{\p_x u_{\scriptscriptstyle l,m}}{u_{\scriptscriptstyle l,m}}\left(u_{\scriptscriptstyle l-1,m} A_{\scriptscriptstyle l,m} P_{\scriptscriptstyle l,m}-\alpha B_{\scriptscriptstyle l,m} Q_{\scriptscriptstyle l,m}\right)+\alpha C_{\scriptscriptstyle l,m} Q_{\scriptscriptstyle l,m}=0,
\]
where $P_{\scriptscriptstyle l,m}$ and $Q_{\scriptscriptstyle l,m}$ are the differences between the left- and right-hand sides of the respective equations in system (\ref{NLS}).
Thus, as expected, relations (\ref{xderivative}) and (\ref{tderivative}) are compatible modulo system~(\ref{NLS}).
\end{Remark}

Once the solution $(u_{\scriptscriptstyle l-1,m+1},u_{\scriptscriptstyle l,m+1})$ is found, we can proceed to construct more involved solutions in a purely algebraic manner.
This step is the standard step in generating the multi-soliton solutions from an auto-B\"acklund transformation and commutative Bianchi diagram.
Assuming that transformations with parameters $\alpha_1$ and $\alpha_2$ commute, i.e., if we repeat the procedure of the previous step twice: first with parameter $\alpha_1$ then~$\alpha_2$, and second, with $\alpha_2$ then $\alpha_1$ -- we get the same result. This simple assumption yields a formula of superposition of solutions for system~(\ref{NLS}). The Bianchi diagram adapted for this calculation is given below:

\begin{center}
\begin{tikzpicture}[thick,scale=0.66, every node/.style={scale=0.66}]
				\node [style=Empty vertex] (0) at (-18, 4) {};
		\node [style=Empty vertex] (1) at (-16.5, 4) {};
		\node [style=Black vertex] (7) at (-16.5, 10) {};
		\node [style=gray node] (18) at (-21, 7) {};
		\node (25) at (-18.25, 3.5) {${ u_{l-1,m,n}}$};
		\node (31) at (-22, 7) {${ u_{l-1,m+1,n}\hspace{2mm}}$};
		\node (32) at (-18.5, 7) {${ u_{l,m+1,n}}$};
		\node (33) at (-16, 3.5) {${ u_{l,m,n}}$};
		\node (34) at (-12.5, 7) {${ u_{l,m,n+1}}$};
		\node [style=Black vertex] (46) at (-18, 10) {};
		\node [style=gray node] (49) at (-15, 7) {};
		\node (50) at (-16.25, 7) {${ u_{l-1,m,n+1}}$};
		\node (51) at (-15.75, 10.5) {${ u_{l,m+1,n+1}}$};
		\node (52) at (-18.5, 10.5) {${ u_{l-1,m+1,n+1}}$};
		\node (53) at (-18.75, 5.5) {};
		\node (54) at (-18.75, 5.5) {$\alpha_1$};
		\node (56) at (-15.75, 5.5) {$\alpha_2$};
		\node (57) at (-18.75, 8.5) {$\alpha_2$};
		\node (58) at (-15.75, 8.5) {$\alpha_1$};
		\node [style=Empty vertex] (59) at (-13.5, 7) {};
		\node [style=Empty vertex] (60) at (-19.5, 7) {};
		\draw (0) to (18);
		\draw [style=plain] (18) to (46);
		\draw [style=new edge style 0] (0) to (1);
		\draw [style=new edge style 0] (46) to (7);
		\draw [style=plain] (0) to (49);
		\draw [style=plain] (46) to (49);
		\draw (1) to (59);
		\draw (59) to (7);
		\draw [style=new edge style 0] (59) to (49);
		\draw (60) to (7);
		\draw [style=new edge style 0] (18) to (60);
		\draw (1) to (60);
\end{tikzpicture}
\end{center}

\noindent Here we have slightly modified notation by introducing an additional discrete variable $n$ which counts auto-B\"acklund transformations with respect to parameter $\alpha_2$.
The respective formula of superposition is obtained by solving a system of four copies of (\ref{lSG}):
\begin{gather*}
\alpha_1(u_{\scriptscriptstyle l-1,m,n}u_{\scriptscriptstyle l,m+1,n}-u_{\scriptscriptstyle l-1,m+1,n}u_{\scriptscriptstyle l,m,n})+u_{\scriptscriptstyle l-1,m+1,n}u_{\scriptscriptstyle l-1,m,n}u_{\scriptscriptstyle l,m,n}u_{\scriptscriptstyle l,m+1,n} = 1, \\
\alpha_2(u_{\scriptscriptstyle l-1,m+1,n}u_{\scriptscriptstyle l,m+1,n+1}-u_{\scriptscriptstyle l-1,m+1,n+1}u_{\scriptscriptstyle l,m+1,n})
 + u_{\scriptscriptstyle l-1,m+1,n+1}u_{\scriptscriptstyle l-1,m+1,n}u_{\scriptscriptstyle l,m+1,n}u_{\scriptscriptstyle l,m+1,n+1} = 1,\\
\alpha_2(u_{\scriptscriptstyle l-1,m,n}u_{\scriptscriptstyle l,m,n+1}-u_{\scriptscriptstyle l-1,m,n+1}u_{\scriptscriptstyle l,m,n})+u_{\scriptscriptstyle l-1,m,n+1}u_{\scriptscriptstyle l-1,m,n}u_{\scriptscriptstyle l,m,n}u_{\scriptscriptstyle l,m,n+1} = 1, \\
\alpha_1(u_{\scriptscriptstyle l-1,m,n+1}u_{\scriptscriptstyle l,m+1,n+1}-u_{\scriptscriptstyle l-1,m+1,n+1}u_{\scriptscriptstyle l,m,n+1})
+u_{\scriptscriptstyle l-1,m+1,n+1}u_{\scriptscriptstyle l-1,m,n+1}u_{\scriptscriptstyle l,m,n+1}u_{\scriptscriptstyle l,m+1,n+1} = 1
\end{gather*}
for $(u_{\scriptscriptstyle l-1,m+1,n+1},u_{\scriptscriptstyle l,m+1,n+1})$. Namely, it is given by
\begin{gather}
	u_{\scriptscriptstyle l-1,m+1,n+1}
	= \frac{\alpha_2^2-\alpha_1^2-u_{\scriptscriptstyle l-1,m,n}\big(\alpha_2 \big(1-\alpha_1^2\big) u_{\scriptscriptstyle l,m,n+1} -\alpha_1\big(1-\alpha_2^2\big) u_{\scriptscriptstyle l,m+1,n}\big)}{
	\big(\alpha_2^2-\alpha_1^2\big)u_{\scriptscriptstyle l,m,n+1} u_{\scriptscriptstyle l-1,m,n} u_{\scriptscriptstyle l,m+1,n}+\alpha_1\big(1-\alpha_2^2\big) u_{\scriptscriptstyle l,m,n+1}
	-\alpha_2\big(1-\alpha_1^2\big)u_{\scriptscriptstyle l,m+1,n}
	},
\nonumber\\
	u_{\scriptscriptstyle l,m+1,n+1} = \frac{ u_{\scriptscriptstyle l,m,n} (\alpha_1 u_{\scriptscriptstyle l,m,n+1} -\alpha_2 u_{\scriptscriptstyle l,m+1,n} ) }{ \alpha_1 u_{\scriptscriptstyle l,m+1,n} -\alpha_2 u_{\scriptscriptstyle l,m,n+1}}.
	\label{spp2}
\end{gather}
The process of constructing solutions by means of the obtained auto-B\"acklund transformation and superposition formula is summarised as follows.\vspace{2mm}

\begin{Summary} Given a seed solution $(u_{\scriptscriptstyle l-1,m},u_{\scriptscriptstyle l,m})$ of (\ref{NLS}) we first solve a system of (\ref{xderivative}) and (\ref{tderivative}) to determine the component $u_{\scriptscriptstyle l,m+1}$.
The other component $u_{\scriptscriptstyle l-1,m+1},$ is calculated algebraically from (\ref{2lSG}). Further, we set
\begin{gather}
\begin{split}
	&(u_{\scriptscriptstyle l-1,m+1,n},u_{\scriptscriptstyle l,m+1,n})=(u_{\scriptscriptstyle l-1,m+1},u_{\scriptscriptstyle l,m+1})|_{\alpha=\alpha_1},\\
	&(u_{\scriptscriptstyle l-1,m,n+1},u_{\scriptscriptstyle l,m,n+1})=(u_{\scriptscriptstyle l-1,m+1},u_{\scriptscriptstyle l,m+1})|_{\alpha=\alpha_2}
\end{split}
	\label{subs2par}
\end{gather}
and substitute these expressions in the superposition formulae~(\ref{spp2}) to calculate the solution $(u_{\scriptscriptstyle l-1,m+1,n+1},u_{\scriptscriptstyle l,m+1,n+1})$.
\end{Summary}

\section{Examples}\label{exactsln}

If we start with a constant solution
\[
u_{\scriptscriptstyle l-1,m}=u_{\scriptscriptstyle l,m}=1,
\]
then equations (\ref{xderivative}) and (\ref{tderivative}) turn into the system
\begin{gather*}
\p_x u_{\scriptscriptstyle l,m+1} = \frac{\alpha}{\alpha^2-1} \big(u_{\scriptscriptstyle l,m+1}^2-1\big), \\
 \p_t u_{\scriptscriptstyle l,m+1} = -\frac{\alpha\big(\alpha^2+1\big)}{\big(\alpha^2-1\big)^2} \big(u_{\scriptscriptstyle l,m+1}^2-1\big),
\end{gather*}
whose solution $u_{\scriptscriptstyle l,m+1}$ can be expressed in terms of the functions
\begin{gather}
	u_{\scriptscriptstyle l,m+1}=f(\alpha,\beta)=\frac{f_+(\alpha,\beta)}{f_-(\alpha,\beta)},
\label{fexpr}
\end{gather}
where
\[
f_{\pm}(\alpha,\beta)=1\pm\beta \exp\left(\frac{2\alpha x}{\alpha^2-1}-\frac{2\alpha\big(\alpha^2+1\big)t}{\big(\alpha^2-1\big)^2}\right).
\]
The other component of the solution is found from (\ref{2lSG}) and given by
\begin{gather}
	u_{\scriptscriptstyle l-1,m+1}=\frac{1-\alpha f(\alpha,\beta)}{f(\alpha,\beta)-\alpha}.	\label{ulmin1mp1}
\end{gather}
Applying substitutions (\ref{subs2par}) to
functions (\ref{fexpr}), (\ref{ulmin1mp1}), and then substituting the resulting expressions in (\ref{2lSG}) we obtain the solution
\begin{gather}
 u_{\scriptscriptstyle l-1,m+1,n+1}=
	\frac{\alpha_1^2-\alpha_2^2-\alpha_1\big(1-\alpha_2^2\big) f(\alpha_1,\beta_1)+\alpha_2\big(1-\alpha_1^2\big) f(\alpha_2,\beta_2)}{\big(\alpha_1^2-\alpha_2^2\big)f(\alpha_1,\beta_1) f(\alpha_2,\beta_2)+\alpha_2\big(1-\alpha_1^2\big) f(\alpha_1,\beta_1)-\alpha_1\big(1-\alpha_2^2\big)f(\alpha_2,\beta_2)},\nonumber\\
	u_{\scriptscriptstyle l,m+1,n+1}=\frac{\alpha_1 f(\alpha_2,\beta_2)-\alpha_2 f(\alpha_1,\beta_1)}{\alpha_1 f(\alpha_1,\beta_1)-\alpha_2 f(\alpha_2,\beta_2)}.
\label{2solit}
\end{gather}
Note that the components $u_{\scriptscriptstyle l-1,m}$ and $u_{\scriptscriptstyle l,m}$ are not conserved densities of equation (\ref{NLS}). Therefore, a quantity of interest is the simplest conserved density given by the formula
\[\rho=(\p_x \ln u_{\scriptscriptstyle l-1,m+1,n+1})(\p_x \ln u_{\scriptscriptstyle l,m+1,n+1}).
\]
For real $\alpha_i$ and imaginary $\beta_i$, this quantity represents an interaction of two solitons:
the profile consists of two distinct solitons with one increasing in amplitude while overtaking
the other and then restoring its shape after the interaction.
Solution~(\ref{2solit}) can be re-written in a more compact form in terms of the skew-symmetric Levi-Chivita symbol $\varepsilon^{ij}$:
\begin{gather*}
u_{\scriptscriptstyle l-1,m+1,n+1}=\frac{-2 \varepsilon^{ij} \alpha_i\big(\alpha_i+\big(\alpha_j^2-1\big)f_i\big) }{\varepsilon^{ij} \big(\alpha_j^2-\alpha_i^2\big)f_i f_j+2 \big(\alpha_i^2-1\big)\alpha_j f_i},
\\ u_{\scriptscriptstyle l,m+1,n+1}=-\frac{\varepsilon^{ij} \alpha_j f_i}{\varepsilon^{ij} \alpha_i f_i},
\end{gather*}
where we assume summation over repeated indices and denote $f_m=f(\alpha_m,\beta_m)$.

{\samepage As a concluding example we construct a three-soliton solution of equation (\ref{NLS}).
The respective stack of Bianchi diagrams (Bianchi lattice) (see, e.g., \cite{schief}) is given by

\begin{center}
\begin{tikzpicture}[thick,scale=0.7, every node/.style={scale=0.7}]
		\node [style=Empty vertex] (0) at (-18, 4) {};
		\node [style=Empty vertex] (1) at (-16.5, 4) {};
		\node (25) at (-18.5, 3.5) {${ u_{l-1,m,n}}$};
		\node (31) at (-21.75, 7.5) {${ u_{l-1,m+1,n}}$};
		\node (32) at (-18.5, 7) {${ u_{l,m+1,n}}$};
		\node (33) at (-16, 3.5) {${ u_{l,m,n}}$};
		\node (34) at (-12.5, 7) {${ u_{l,m,n+1}}$};
		\node (50) at (-16.25, 7) {${ u_{l-1,m,n+1}}$};
		\node (51) at (-15.25, 10) {${ u_{l,m+1,n+1}}$};
		\node (52) at (-19.25, 10) {${ u_{l-1,m+1,n+1}}$};
		\node [style=Empty vertex] (53) at (-12, 4) {};
		\node [style=Empty vertex] (54) at (-10.5, 4) {};
		\node [style=Black vertex] (57) at (-15, 13) {};
		\node [style=Black vertex] (58) at (-13.5, 13) {};
		\node (63) at (-18.75, 5.5) {$\alpha_1$};
		\node (64) at (-15.75, 8.25) {$\alpha_1$};
		\node (65) at (-12.5, 11.5) {$\alpha_1$};
		\node (66) at (-15.75, 5.5) {$\alpha_2$};
		\node (67) at (-19, 8.25) {$\alpha_2$};
		\node (68) at (-13, 8.25) {$\alpha_3$};
		\node (69) at (-16, 11.25) {$\alpha_3$};
		\node (70) at (-9.75, 5.5) {$\alpha_3$};
		\node (71) at (-12.75, 5.5) {$\alpha_2$};
		\node (72) at (-9.5, 8.25) {$\alpha_2$};
		\node [style=Empty vertex] (74) at (-19.5, 7) {};
		\node [style=Empty vertex] (75) at (-13.5, 7) {};
		\node [style=Empty vertex] (76) at (-7.5, 7) {};
		\node (83) at (-12.5, 3.5) {${ u_{l-1,m,n}}$};
		\node (84) at (-10, 3.5) {${ u_{l,m,n}}$};
		\node (85) at (-10, 7) {${ \hat{u}_{l-1,m,n}}$};
		\node (86) at (-7, 7.5) {${ \hat{u}_{l,m,n}}$};
		\node [style=Empty vertex] (89) at (-15, 7) {};
		\node [style=Empty vertex] (90) at (-16.5, 10) {};
		\node [style=Empty vertex] (91) at (-10.5, 10) {};
		\node (92) at (-13, 10) {${ \hat{u}_{l-1,m,n+1}\hspace{3mm}}$};
		\node (93) at (-9.5, 10) {${ \hat{u}_{l,m,n+1}}$};
		\node (94) at (-15.75, 13.5) {${ \hat{u}_{l-1,m+1,n+1}}$};
		\node (95) at (-12.75, 13.5) {${ \hat{u}_{l,m+1,n+1}}$};
		\node [style=Empty vertex] (96) at (-21, 7) {};
		\node [style=Empty vertex] (97) at (-9, 7) {};
		\node [style=Empty vertex] (98) at (-18, 10) {};
		\node [style=Empty vertex] (99) at (-12, 10) {};
 \draw [style=new edge style 0] (0) to (1);
		\draw [style=new edge style 0] (53) to (54);
		\draw [style=new edge style 0] (57) to (58);
		\draw (1) to (74);
		\draw (1) to (75);
		\draw [style=plain] (54) to (75);
		\draw [style=plain] (54) to (76);
		\draw [style=plain] (0) to (89);
		\draw [style=plain] (53) to (89);
		\draw [style=new edge style 0] (75) to (89);
		\draw (90) to (58);
		\draw (74) to (90);
		\draw (75) to (90);
		\draw [style=plain] (91) to (58);
		\draw [style=plain] (75) to (91);
		\draw [style=plain] (76) to (91);
		\draw [style=plain] (53) to (97);
		\draw [style=new edge style 0] (97) to (76);
		\draw (0) to (96);
		\draw [style=new edge style 0] (96) to (74);
		\draw (98) to (57);
		\draw [style=plain] (98) to (89);
		\draw [style=new edge style 0] (98) to (90);
		\draw [style=plain] (96) to (98);
		\draw [style=plain] (99) to (57);
		\draw [style=plain] (89) to (99);
		\draw [style=new edge style 0] (99) to (91);
		\draw [style=plain] (97) to (99);
\end{tikzpicture}
\end{center}

}

\noindent Here the result of the application of a B\"acklund transformation with parameter $\alpha_3$ to the solution $(u_{l-1,m,n},u_{l,m,n})$
is denoted as $(\hat{u}_{l-1,m,n},\hat{u}_{l,m,n})$.
The diagram gives a three-soliton solution in the form
\begin{gather*}
\hat{u}_{\scriptscriptstyle l-1,m+1,n+1}=\frac{\varepsilon^{ijk}\alpha_i \big(\alpha_j\big(\alpha_i^2-\alpha_j^2\big)\big(\alpha_k^2-1\big)f_j-\big(\alpha_i^2-1\big)\big(\alpha_k^2-\alpha_j^2\big)\big)f_i}{
	\varepsilon^{ijk} \alpha_k \big(\big(\alpha_i^2-\alpha_j^2\big) \big(\alpha_k^2-1\big)f_j-\alpha_j \big(\alpha_i^2-1\big) \big(\alpha_k^2-\alpha_j^2\big)\big)f_i
	}, \\
\hat{u}_{\scriptscriptstyle l,m+1,n+1}=\frac{\varepsilon^{ijk} \alpha_k \big(\alpha_i^2-\alpha_j^2\big)f_if_j}{\varepsilon^{ijk} \alpha_i \big(\alpha_j^2-\alpha_k^2\big) f_i},
\end{gather*}
where $i,j,k=1,\dots,3$.

\begin{figure}[ht]
\centering
 \includegraphics[angle=0,scale=0.6]{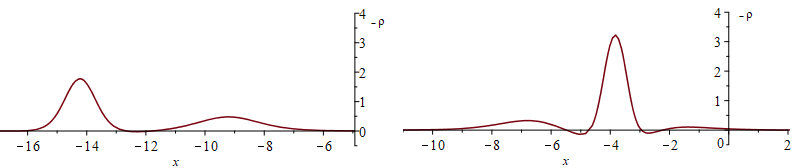} 

\centering
 \includegraphics[angle=0,scale=0.6]{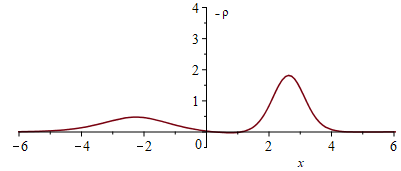} 

 \caption{Interaction of two solitons \big($\alpha_1=2$, $\alpha_2=3$, $\beta_1=\beta_2=5\sqrt{-1}$\big).}
 \end{figure}

As we previously pointed out, the term {\it soliton} solution applies to the conserved density
\[\rho=(\p_x \ln \hat{u}_{\scriptscriptstyle l-1,m+1,n+1})(\p_x \ln \hat{u}_{\scriptscriptstyle l,m+1,n+1})\]
rather than individual components $\hat{u}_{\scriptscriptstyle l-1,m+1,n+1}$ and $\ln \hat{u}_{\scriptscriptstyle l,m+1,n+1}$.
The graphs of $-\rho$ for the values of time $t=-6.5$, $t=-0.5,$ and $t=3$ are as follows

\begin{figure}[ht]
\centering
\includegraphics[angle=0,scale=0.7]{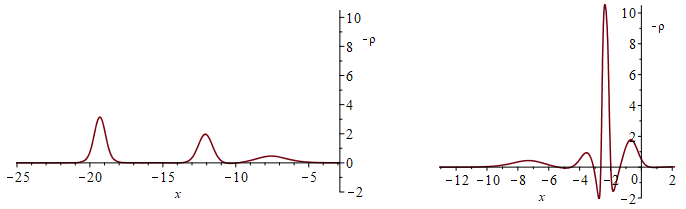} 

 \includegraphics[angle=0,scale=0.7]{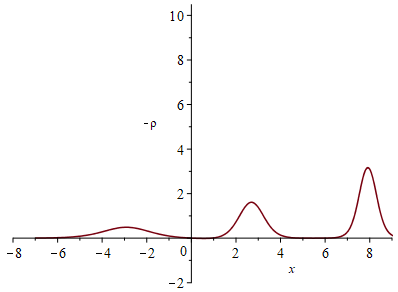} 

 \caption{Interaction of three solitons ($\alpha_1=3/2$, $\alpha_2=2$, $\alpha_3=3$, $\beta_1=\beta_2=\beta_3=7{\rm i}$).}
 \end{figure}

\section{Conclusion}
In this article we have looked at implications of treating the lattice sine-Gordon equation and two of its generalised symmetries as a compatible system. This viewpoint yields, almost automatically, the auto-B\"acklund transformation and algebraic superposition formula for a modified derivative NLS system. The efficacy of the formulae has been verified by constructing two and three soliton solutions.

\pdfbookmark[1]{References}{ref}
\LastPageEnding

\end{document}